\input harvmac
%\draftmode
\def \OO {{\cal O}}
\def \tr {{\rm tr}}
\def \oo {O_8}
\def \o {O_4}
\def \Str {{\rm Str}}
\def \gym{g_{\rm YM} }

\def \D {{\Delta}}

\def \a {\alpha}

\def \four {{1\ov 4}}

\def \ep{\epsilon}
\def \N {{\cal N}}

\def \k {\kappa} 
\def \cF {{\cal F}}

\def \del {\partial}

\def \r {\rho}

\def \p {\phi}
\def \m {\mu}
\def \n {\nu}
\def \vp {\varphi }

\def \Ga {{\Gamma}}

\def \d {\delta}

\def \om {\omega}

\def \ov {\over }
\def \four{{\textstyle{1\over 4}}}

\def \ff {{f^{(2)}_I}}

\def \ha {{ {1 \over 2}}}

\def \G {\Gamma}

\def \l {\lambda}

\def \lr { \lref}
\def\np {{  Nucl. Phys. }}
\def \pl {{  Phys. Lett. }}

%%%%%%%%%%%%%%%%%%%%%%%%%%%%%%%%%%%%%%%%%%%%%

\lr \gks{S.S. Gubser, I.R. Klebanov  and A.A. Tseytlin, 
``Coupling Constant Dependence in the Thermodynamics of $\N=4 $
Supersymmetric Yang-Mills Theory'', hep-th/9805156. }

\lr \freed{D.Z. Freedman, S.D. Mathur, A. Matusis and L. Rastelli,
``Correlation functions in the CFT(d)/AdS(d+1) correspondence'', 
Nucl. Phys. B546 (1999) 96, hep-th/9804058.}

\lr\roch{F. Gonzalez-Rey, B. Kulik, I.Y. Park  and  M. Ro\v cek,
``Self-Dual Effective Action of $\N=4$ Super-Yang Mills",
 hep-th/9810152.}

\lr\kett{S.V. Ketov, 
   ``A manifestly $\N=2$ supersymmetric Born-Infeld action", 
hep-th/9809121.}

\lr\gonr{F. Gonzalez-Rey  and  M. Ro\v cek,
``Nonholomorphic $\N=2$ terms in $\N=4$ SYM: 1-Loop Calculation in 
$\N=2$ superspace", 
Phys.Lett. B434 (1998) 303, 
hep-th/9804010.}

%%%%%%%%%%%%%%%%%%%%%%%%%%%%%%%%%%%%%%%%%%%%%%%%%%%%%%%%%%

\lr \pol{ A.M. Polyakov, ``String theory and
quark confinement", hep-th/9711002; 
 ``A few projects in
string theory", Les Houches Summer  School,  1992:783, 
hep-th/9304146. }

\lr \mal{ J. Maldacena, ``The large $N$ limit of
superconformal
field theories and supergravity", Adv. Theor. Math. Phys. 2 (1998) 231,
{
hep-th/9711200}.}

\lr \gkp {S.S. Gubser, I.R.  Klebanov and A.M.  Polyakov,
``Gauge theory correlators from non-critical string theory",
\pl B428 (1998) 105, 
hep-th/9802109.}

\lr \witt {E.  Witten, ``Anti de Sitter space and holography", 
Adv. Theor. Math. Phys. 2 (1998) 253,
hep-th/9802150.}

\lr \ferr{ S. Ferrara,   C. Fronsdal and A. Zaffaroni, 
``On N=8 Supergravity on $AdS_5$ and $\N=4$
Superconformal Yang-Mills theory", Nucl. Phys. B532 (1998) 153, 
hep-th/9802203;
 L. Andrianopoli  and S. Ferrara,
``K-K excitations on $AdS_5 \times S^5$ as N=4 ``primary'' superfields'', 
Phys. Lett. B430 (1998) 248, 
hep-th/9803171;
S. Ferrara, M. A. Lledo and A. Zaffaroni, 
 ``Born-Infeld Corrections to D3 brane Action in $AdS_5\times S_5$ and $\N=4$, d=4
     Primary Superfields'', 
 Phys. Rev. D58 (1998) 105029, 
hep-th/9805082.}
 
 \lr \schwa{J. Schwarz,   
 ``Covariant field equations of chiral 
 N=2 D=10 supergravity'',
  Nucl. Phys. B226 (1983) 269.}

\lr \kleb { I.R. Klebanov, ``World-volume approach
to absorption by non-dilatonic branes",
\np B496 (1997) 231, hep-th/9702076.}
\lr\klebb{
S.S. Gubser, I.R.  Klebanov and A.A.
  Tseytlin, ``String theory and classical
  absorption by three-branes", 
  \np B499 (1997) 41, hep-th/9703040.}

\lr \gubs{
S.S. Gubser and I.R. Klebanov,``Absorption by branes and Schwinger
terms in the world-volume theory", Phys. Lett. B413 (1997) 41, 
hep-th/9708005.}

\lr\kle{S.S. Gubser, A. Hashimoto, I.R. Klebanov  and  M.
       Krasnitz,
       ``Scalar absorption and the breaking of the world volume conformal
       invariance", Nucl. Phys. B526  (1998) 393, 
       hep-th/9803023;
 S.S. Gubser and A. Hashimoto, ``Exact absorption probabilities for the D3-brane'',
Commun.Math.Phys. 203 (1999) 325, 
hep-th/9805140.}

\lr \volo{I.Ya. Aref'eva and  I.V. Volovich,
   ``On large $N$ conformal theories, field theories in Anti de Sitter 
space  and singletons", hep-th/9803028;
``On the Breaking of Conformal Symmetry in the AdS/CFT Correspondence", hep-th/9804182.}

\lr \aru{ G. Arutyunov  and  S. Frolov,
``Three-point Green function of the stress-energy tensor in the AdS/CFT
      correspondence",  hep-th/9901121.}

\lr \us{H. Liu and A.A. Tseytlin, ``D=4 Super-Yang-Mills, D=5 Gauged
Supergravity, and D=4 Conformal Supergravity,'' 
\np B553 (1998) 88, hep-th/9804083.}

\lr\muck{
W.  M\" uck and K.S. Viswanathan, 
``Conformal field theory correlators from classical scalar field theory on AdS$_{d+1}$", 
hep-th/9804035;
 ``Conformal Field
Theory Correlators from Classical Field Theory on Anti-de Sitter Space II.
Vector and Spinor Fields,'' hep-th/9805145.}

\lr \rchalmers{G. Chalmers, H. Nastase, K. Schalm and R. Siebelink, 
 ``R-Current Correlators in $\N=4$ SYM from $AdS$,''
hep-th/9805015.}
\lr \sei{S. Lee, S. Minwalla, M. Rangamani and N. Seiberg, 
``Three-Point Functions of Chiral Operators in D=4, $\N=4$ SYM at Large N'',
Adv.Theor.Math.Phys.2 (1998) 697, 
hep-th/9806074.}

\lr \fro{G.E. Arutyunov and S. Frolov, ``On the origin of supergravity boundary
terms in the AdS/CFT correspondence'', hep-th/9806216.} 

\lr \howest{G.G. Harwell  and P.S. Howe, ``A superspace survey",
Class. Quant. Grav. 12 (1995) 1823;
P.S. Howe and P.C.  West,
``Non-perturbative Green's functions in theories with
extended superconformal symmetry," 
hep-th/9509140; 
 ``Operator product expansions in 
four-dimensional superconformal field theories'', 
Phys. Lett. B389 (1996) 273, hep-th/9607060.}

\lr \anselm{D. Anselmi, M. Grisaru and A. Johansen, ``A Critical Behaviour 
of Anomalous Currents, Electric-Magnetic Universality and $CFT_4$'', 
hep-th/9601023,
Nucl.Phys. B491 (1997) 221.}

\lr \ansel{D. Anselmi, D. Freedman, 
M. Grisaru and A. Johansen, ``Non-perturbative
formulas for central functions  of supersymmetric gauge theories'', 
Nucl. Phys. B526 (1998) 543, hep-th/9708042.}

%%%%%%%%%%%%%%%%%%%%%%%%%%%%%%%%%%%%%%%%%%%%%%%%%%%
\lr \howes {P.S. Howe, E. Sokatchev and P.C. West,
``3-Point Functions in $\N=4$ Yang-Mills'', 
Phys. Lett. B444 (1998) 341, hep-th/9808162.}

\lr \eden{B. Eden, P.S. Howe and P.C. West, `` Nilpotent invariants in $\N=4$ SYM'',
hep-th/9905085.}

\lr \guys { O. Aharony, S. S. Gubser, J. Maldacena, H. Ooguri  and  Y. Oz,
``Large N Field Theories, String Theory and Gravity'', 
hep-th/9905111.}

\lr \rrey{F. Gonzalez-Rey, B. Kulik  and  I.Y. Park, 
``Non-renormalization of two and three Point Correlators of $\N=4$ SYM in N=1
Superspace'', hep-th/9903094.}

\lr \intri{K. Intriligator, ``Bonus Symmetries of $\N=4$ Super-Yang-Mills 
Correlation Functions via AdS Duality'', hep-th/9811047;
K. Intriligator and W. Skiba, ``Bonus Symmetry and the Operator Product 
Expansion of $\N=4$ Super-Yang-Mills'', hep-th/9905020.}

\lr\ski{ E. D'Hoker, D.Z. Freedman  and  W. Skiba, 
``Field theory tests for correlators in the AdS / CFT correspondence'',  
 Phys. Rev. D59  (1999) 045008,  hep-th/9807098.}

\lr \allrefs{we need a systematic list of refs to 3-point correlators in AdS--
 like in 9905130 }

\lr \dastr  { S.R. Das and   S.T. Trivedi,
``Three-Brane action and the correspondence between $\N=4$ Yang-Mills theory and
 Anti-de Sitter space", \pl B445 (1998) 142, hep-th/9804149;
S. de Alwis, ``Supergravity, the DBI action and the correspondence between the $\N=4$ Yang-Mliis theory and the anti de Sitter space'',
\pl B435 (1998) 31, hep-th/9804019.
 }
\lr \niew { H. J. Kim, L. J. Romans, and P. van Nieuwenhuizen,
``The Mass Spectrum Of Chiral $N=2$ $D=10$ Supergravity on $S^5$'',
Phys. Rev. D32 (1985) 389.}

\lr\gsw { Grimm Sohnius wess}

\lr \leigh{
 R.G. Leigh, Dirac-Born-Infeld action from Dirichlet sigma model,
 Mod. Phys. Lett. A4 (1989) 2767.}

\lr \mets { R.R. Metsaev  and M.A. Rakhmanov,
``Fermionic terms in the open superstring effective action'',
\pl B193 (1987) 202;
 R.R. Metsaev, M.A. Rakhmanov  and  A.A. Tseytlin, ``The Born-Infeld
     action as the effective action in the open superstring theory, 
    Phys. Lett. B193 (1987) 207.
 }

\lr \frad{E.S. Fradkin  and A.A.  Tseytlin, ``Non-linear electrodynamics from quantized strings",
\pl B163 (1985) 123. }
\lr\napp{   A. Abouelsaood, C.G. Callan, C.R. Nappi  and  S.A. Yost,
``Open strings in background fields", 
Nucl. Phys. B280 (1987) 599. } 

\lr \berg{ E. Bergshoeff, M. Rakowski, E. Sezgin, Higher derivative 
    super Yang-Mills theories, 
    Phys. Lett. B185  (1987)  371.
}

\lr\sch{ 
%M. Cederwall, A. von Gussich, B.E.W. Nilsson and A. Westenberg,
%The Dirichlet super-three-brane in ten-dimensional type IIB supergravity,
%Nucl. Phys. {B}490 (1997) 163, hep-th/9610148; 
M. Aganagic, C. Popescu and J.H. Schwarz, D-brane actions with local kappa-symmetry,
Phys. Lett. {B}393 (1997) 311, hep-th/9610249;
Gauge-invariant and and gauge-fixed D-brane actions,
Nucl. Phys. {B}495 (1997) 99, hep-th/9612080; 
%E. Bergshoeff and  P.K. Townsend, Super D-branes,
%Nucl. Phys. {B}490 (1997) 145,
%hep-th/9611173.
}

\lr \disc {A. Hashimoto  and  I.R. Klebanov, 
  ``Decay of excited D-branes'', 
Phys. Lett. B381  (1996) 437-445, hep-th/9604065;  
M.R. Garousi   and  R.C. Myers, 
 World volume intercations on D-branes,  
 Nucl. Phys. B542  (1999) 73, 
hep-th/9809100.   } 

\lr\roche{  F. Gonzalez-Rey, I.Y. Park  and  M. Rocek, 
``On dual 3-brane actions with partially broken N=2 supersymmetry'',
 Nucl. Phys. B544  (1999) 243, 
 hep-th/9811130.
   } 

\lr \GSW { Green Scwarz Witten book}

\lr \dis  {Dine Seiberg }
%\lr \roch {Rochek et al about $F^4$ term in 1-loop action }
\lr \buchb  {Buchbinder et al about $F^4$ term in 1-loop action }

\lr \bagg { Bagger and Galperin  \pl B336 (1994) 25}

\lr \rot  { M. Rocek and A.A. Tseytlin, ``Partial breaking of global 
D=4 supersymmetry, constrained superfields, and 3-brane actions'',
Phys.Rev. D59 (1999) 106001,  hep-th/9811232.}

\lr \banks { T. Banks  and M.B.  Green,
  ``Nonperturbative effects in  $  AdS_5 \times S^5 $ 
string theory and d = 4 SUSY
Yang-Mills",  
 JHEP 9805  (1998) 002, 
 hep-th/9804170.
 } 

\lr \ktv{ I. Klebanov, W. Taylor  and  M. Van Raamsdonk,
``Absorption of dilaton partial waves by D3-branes'', 
hep-th/9905174.}

\lr \LT {H. Liu and  A.A. Tseytlin, ``On Four-point Functions 
in the CFT/AdS Correspondence'',
Phys. Rev. D59 (1999) 086002, hep-th/9807097.}

\lr \fref {D.Z. Freedman, S.D. Mathur, A. Matusis and L. Rastelli,
``Comments on 4-point functions in the CFT/AdS correspondence'',
Phys. Lett. B452 (1999) 61, hep-th/9808006}

\lr \fres {E. D'Hoker, D.Z. Freedman, S.D. Mathur, A. Matusis and L. Rastelli,
``Graviton exchange and complete 4-point functions in the AdS/CFT 
correspondence'', hep-th/9903196.}

\lr \Witten {E. Witten, unpublished.}

\lr \feraz{ Ferrara BI action }

\lr \liu {H. Liu, ``Scattering in Anti-de Sitter Space and Operator 
Product Expansion'', hep-th/9811152.}

\lr \dhfr {E. D'Hoker and D. Z. Freedman, 
``General Scalar Exchange in AdS$_{d+1}$'',
hep-th/9811257.}

\lr \chal {G. Chalmers and K. Schalm, ``The large $N_c$ limit of 
four-point functions in $\N=4$ super-Yang-Mills theory from
anti-de Sitter Supergravity'', hep-th/9810051}

\lr \chalf {G. Chalmers, H. Nastase, K. Schalm  and  R. Siebelink, ``R-Current
 Correlators in $\N=4$ super Yang-Mills theory from Anti-de Sitter
supergravity'', Nucl.Phys. B540 (1999) 247, hep-th/9805105.}

\lr \call {C.G. Callan, S.S. Gubser, I.R. Klebanov and A.A. Tseytlin,
``Absorption of Fixed scalars and the D-brane Approach to Black Holes'',
Nucl. Phys. B489 (1997) 65, hep-th/9610172.}

\lr \pet {A. Petkou and K. Skenderis, ``A non-renormalization theorem for conformal anomalies", 
hep-th/9906030.}

\lr \grey {F. Gonzalez-Rey, H. Liu and  A.A. Tseytlin, to appear.}

\lr \tsea{  A.A. Tseytlin, 
``Selfduality of Born-Infeld action and Dirichlet three-brane of type IIB
 superstring theory'', Nucl. Phys. B469 (1996) 51,  hep-th/9602064. 
 } 

%%%%%%%%%%%%%%%%%%%%%%%%%%%%%%%%%%%%%
\baselineskip12pt
\Title{
\vbox
{\baselineskip 6pt{\hbox{   }}{\hbox
{Imperial/TP/98-99/56}}
{\hbox
{NSF-ITP-99-060}}
{\hbox{hep-th/9906151}} {\hbox{
  }}} }
{\vbox{\centerline {
Dilaton -- fixed scalar correlators }
%and AdS/CFT correspondence}
%{\vbox{\centerline {  } 
\vskip4pt
 \centerline { and  $AdS_5 \times S^5$  --  SYM  correspondence} 
%\vskip4pt
 %\centerline {     }
}}
%
%%%%%%%%%%%%%%%%%%%%%%%%%%%%%%%%%%%%%%
\vskip -32 true pt
\bigskip
%\centerline{ F. Gonzalez-Rey\footnote{$^\natural$}{\baselineskip8pt
%e-mail address:  glezrey@pha.jhu.edu}}

%\smallskip
%\centerline{\it Department of Physics and Astronomy,  }
%\smallskip
%\centerline {\it  The  Johns Hopkins University, Baltimore MD 21218, USA }

\bigskip

\centerline{ Hong Liu$^1$\footnote{$^\sharp$}{\baselineskip8pt
e-mail address: hong.liu@ic.ac.uk} and 
A.A. Tseytlin$^{1,2}$\footnote{$^{\star}$}{\baselineskip8pt
e-mail address: tseytlin@ic.ac.uk}\footnote{$^{\dagger}$}
{\baselineskip8pt Also at  Lebedev  Physics
Institute, Moscow.}}

\medskip
 \centerline {$^1$\it  Theoretical Physics Group, Blackett Laboratory,}
\smallskip
\centerline {\it  Imperial College,  London SW7 2BZ, U.K. }

\medskip
\centerline {$^2$\it Institute   for    Theoretical Physics, University of
California,}
\smallskip
\centerline {\it  Santa Barbara, CA93106, USA}

\bigskip
\centerline {\bf Abstract}
\medskip
\medskip
\baselineskip12pt
\noindent
We  address the question of AdS/CFT correspondence 
in the case of  the  3-point function $<O_4O_4O_8>$.  
 $O_4$ and $O_8$ are particular
 primary states represented by $ \tr  F^2 + ...$ and  $ \tr  F^4 + ...$
operators in $\N=4$  SYM theory   and    dilaton  $\p$  and   massive 
 ``fixed'' scalar $\n$  in  $D=5$ supergravity. 
While the  value of  $<O_4O_4O_8>$ 
computed in large N weakly coupled SYM  theory  is  non-vanishing, 
the $D=5$ action of type IIB supergravity compactified on $S^5$ 
 does not contain 
$\phi\phi\nu$ coupling and thus the corresponding 
correlator  seems to vanish on the $AdS_5$ side. 
This is in obvious contradiction  with 
various arguments suggesting    non-renormalization of  2- and 3-point 
functions of states from short multiplets and    implying  
agreement between the supergravity and SYM expressions for them.
We propose  a  natural resolution  of  this paradox
which emphasizes the 10-dimensional nature of the correspondence.
The basic idea is to  treat  the  constant  mode of the dilaton 
as a  part of the full $S^5$ Kaluza-Klein family of dilaton modes. 
This leads to a  non-zero  result for the $<O_4O_4O_8>$ 
correlator on the supergravity side.

%In the process, we  present several new  technical results. 
%In particular, we the $\N=1$ and $\N=2$  manifestly supersymmetric   
%form of the $\oo$ operator  and develop 
%the $\N=2$ projective superspace  formalism  for computing its correlators
%in the free-field limit. The structure
%of the scalar $ (\del \Phi)^4$ part of the $\oo$ operator  which follows 
%  from the Born-Infeld  D3-brane action 
%is shown to be consistent with extended supersymmetry. 

\bigskip
\vskip 20 true pt

%%%%%%%%%%%%%%%%%%%%%%%%%%%%%%%%%%%%%%%%%%%%%%%%%%%%%%%%%
\Date {June 1999 }
%%%%%%%%%%%%%%%%%%%%%%%%%%%%%%%%%%%%%%%%%%%%%%%%%%%%%%%%%%%%%%%%%%%
\noblackbox
\baselineskip 14pt plus 2pt minus 2pt
%\baselineskip 20pt plus 2pt minus 2pt
%%%%%%%%%%%%%%%%%%%%%%%%%%%%%%%%%%%%%%%%%%%

%%%%%%%%%%%%%%%%%%%%%%%%%%%%%%
\newsec{Introduction}
%%%%%%%%%%%%%%%%%%%%%%%%%%%%%%%%

Recent studies of  2- and 3-point functions of operators in short multiplets 
of $\N=4$ super Yang-Mills theory  in the context of   AdS/CFT correspondence 
\refs{\mal,\gkp,\witt}\ (see \guys\ for a 
review) suggest that they are  not renormalized in large $N$ limit 
\refs{\gubs,\gkp,\freed,\banks,\us,\chalf,\sei,\aru,\ktv}.\foot{For other papers  on correlators 
in the AdS/CFT correspondence 
see \refs{\volo,\muck}  and references  in  \guys.}
Since the supergravity correlator is interpreted as the 
large $N$ strong coupling SYM result, 
this implies   that the exact SYM  expressions 
for the  large $N$ limit of the 
2- and 3-point correlators  of such operators 
  should  not contain   non-trivial  functions  of `t Hooft coupling
(the coordinate dependence
of the $n\leq 3$ correlators 
 is fixed uniquely by conformal invariance). 
 This  conclusion  is 
 supported by
 explicit SYM perturbative 
calculations \refs{\ski,\howes,\eden,\rrey},  
 general arguments based on $\N=4$
superspace \eden\  or   $U(1)_Y$ symmetry \intri\  
(or non-renormalization of anomalies \pet).
%and  is also consistent
%with expected non-renormalization
%of 2- and 3- point functions 
%in  string theory  in a maximally supersymmetric background 
%\banks.
However,   one easily  finds an apparent contradiction.

Consider the following $SU(N)$  $\N=4$ SYM operators 
\eqn\one{
\o = \tr(F_{\m \n}  F_{\m \n}) + ... \ ,  }
$$
\oo
= { 3 \ov 2} {\rm Str} [ F^4 - { 1 \ov 4} (F^2)^2 ]  + ...
$$
\eqn\two{
=  \tr (F_{\m \n} F_{\r \n} F_{\m \l} F_{\r \l}
 + \ha  F_{\m \n} F_{\r \n} F_{\r \l} F_{\m \l}
- {1 \ov 4} F_{\m \n} F_{\m \n} F_{\r \l} F_{\r \l}
- {1 \ov 8} F_{\m \n} F_{\r \l} F_{\m \n} F_{\r \l}) + ...\ , 
} 
where dots stand for scalar and spinor terms 
which are required by $\N=4$ supersymmetry
%complete  the gauge field structures  $\N=4$ superinvariants
and Str is the symmetrized trace of $SU(N)$ generators
as matrices in the fundamental representation. 
$O_4$ and $O_8$ are supersymmetric descendants of chiral primary 
operators $\tr (XX)$ and $\tr (XXXX)$, respectively,\foot{The precise definition 
for these chiral primary operators is \sei: 
$C^I_{i_1 \cdots i_k} \tr (X^{i_1} \cdots X^{i_k})$,
where $C^I$ is a totally symmetric traceless rank $k$ tensor  of $SO(6)$
and $X^i$ are scalars of $\N=4$  SYM theory.
%$TrX^{(i} X^{j)}$ and $TrX^{(i_1}X^{i_2}X^{i_3}X^{i_4)}$ 
The action  of supercharges in $\tr X^2$ and $\tr X^4$ 
 is indicated only schematically. 
In particular, in  $O_8$  one needs to  specify the ordering  of $Q$'s and $\bar Q$'s   
to resolve the associated  total derivative 
ambiguity.} 
\eqn\reew{
O_4 = ( Q^4 + \bar{Q}^4)   \tr X^2\ ,
%$TrX^{(i} X^{j)}$, 
\ \,\,\ \,\,\ \,\,\ \,\,
O_8 = Q^4 \bar{Q}^4 \tr X^4 \ . 
%$TrX^{(i_1}X^{i_2}X^{i_3}X^{i_4)}$  \ .
}
For  all the operators $\tr X^n$  the precise 
agreement between 
the (weighted)  3-point correlators computed in the 
large $N$ weakly coupled SYM and the $AdS_5$ supergravity
  was demonstrated in   
\sei, so it  seems natural to expect the same 
for their supersymmetry descendants.  

Using  the  AdS/CFT  
correspondence recipe, one  finds -- either from the 
supermultiplet considerations \refs{\witt,\ferr} or 
from the structure of the  Born-Infeld  action for 
a  D3-brane in  a curved background \tsea\ --  that on the
 supergravity side
$O_4$ corresponds  to the 
5-D dilaton $\p$ \refs{\kleb,\klebb}, while $O_8$ corresponds  
 to the fixed scalar $\n$ \kle.\foot{A relation between 
 fixed scalars and  $F^4$-type operators 
was earlier mentioned in \call\ in the case of the D5+D1 system.} 
The latter  enters the $D=10$  Einstein-frame metric as
\eqn\radi{
ds^2_{10E} = e^{-{10 \ov 3} \n(x)} g_{5 mn} (x) dx^m dx^n + 
e^{2 \n(x)} d \Omega_5^2
\ .  } 
The specific powers of $e^\n$  in the metric
are needed to decouple $\n$ from the 5-d graviton;
this ansatz generalizes to non-linear level
the graviton mode  decomposition  considered in  \niew\ where 
this  massive ($m^2 = { 32 \ov R^2}$) singlet 
scalar  $\n$ was 
identified with  the zero mode ($S^5$ independent part) of the trace
of the  perturbation of the metric of $S^5$.  
%from its $AdS_5$ background value. 
Note that the zero mode $\nu$  
does not mix with the 4-form potential, while
higher Kaluza-Klein  modes do \niew. It is the lowest-mass member  of the KK family
of scalars with masses $m^2 = { (k+4)(k+8) \ov R^2}, \ k=0,1,...$,
and belongs to  a separate  massive  $D=5$  supermultiplet  interacting 
with  the $D=5,\ \N=8$ supergravity multiplet.  
Dimensionally reducing the dilaton-graviton sector of 
type IIB supergravity action to 5 dimensions, one easily 
finds \LT\   
that the action does {\it not} contain  $\phi\phi\nu$ coupling
(in agreement with  
 the possibility to perform the so-called consistent 
truncation of the full $D=5$  theory 
 to the $D=5$  gauged supergravity subsector). 
This implies  that  on the  AdS supergravity side 
$$
 <O_4 O_4 O_8>_{\rm AdS}\ = \  0 .
$$
However, the   explicit computation of $<O_4 O_4 O_8>$ 
 in 
the  SYM   theory 
 performed below to 
the leading order in large $N$  expansion 
%(which for fixed 't Hooft coupling means also   to leading order in $\gym$ 
%or in the free theory limit)
yields a  non-vanishing  result  for this correlator.\foot{That 
this  correlator of the bosonic gauge-theory  parts
 of $O_4$ and $O_8$ computed in the abelian theory limit 
is non-vanishing was already observed  in  the first paper in 
\kle.}

 One  then faces the following   three options:

(i) There  may be  some  subtlety
 in applying the nonrenormalization theorem 
for the   3-point functions of  chiral primary operators $\tr X^n$ 
 to their supersymmetry descendants and the  correlator $<O_4 O_4 O_8>$  may
actually depend on gauge coupling in such a way  that  
it vanishes in the limit $\l\equiv \gym^2 N \to \infty $.\foot{One possibility
 could be  that the $\N=4$  supersymmetric expression for the 3-point function for 
the whole multiplet  contains several tensor structures. However,  
this  seems to be excluded by the arguments  in 
\refs{\howes,\eden}.}
 
(ii) The non-renormalization statement for  the SYM correlator
is correct but there may be  a  subtlety   in the definition of the 
$O_8$ operator, e.g.,  it may contain  also 
${ 1 \ov N} \tr F^2 \tr F^2$ 
admixture\foot{We thank  I. Klebanov and J. Polchinski for suggesting 
this possibility.} (with various possible Lorentz contractions of the two $F^2$ factors). That  would  change the result for  the three-point
function on  the SYM side  and may make it zero.\foot{The
 leading large $N$  term in  $<(\tr F^2)^4 >$ which   is of order $O(1)$
 comes  from
disconnected diagram, while the next order term (leading connected
contribution) is of order $ O(1/N^2)$. One can show (using the results 
of section 2 below) that 
 $<O_4 O_4 O'_8>=0$ if $O'_8=O_8 -  { 1 \ov 4 N} \tr F^2 \tr F^2$.}
% The contribution of ${ 1 \ov N} \tr F^2 \tr F^2$ 
%to F^4 should come from disconnected diagrams.
However, this suggestion seems to be in contradiction with arguments based on $\N=4$ supersymmetry \refs{\howes,\howest}.

(iii) There may be  subtleties in the supergravity calculation.
The  supergravity result may turn out to be non-vanishing 
and in agreement with the non-vanishing  large $N$ 
 SYM result.

%While there is no direct proof of non-renormalization of 
%$<O_8O_8>$  and $<O_4O_4O_8>$  correlators 
%this is highly plausible, so the resolution should be on supergravity side. 

In this paper we shall  propose a  natural resolution of this paradox
based on  the third option.
The key observation is the following. 
Suppose that there was 
some  non-vanishing
 two dilaton -- fixed scalar  coupling in
the $D=5$ action. 
Then  the resulting 3-point function computed from supergravity
according to the rules of \witt\ 
would   actually be  {\it divergent}.
 This can be seen from the general expression for the 
3-point function of  the  scalar fields in the (Euclidean)  $AdS_5$ 
space found  in \freed.
If the scalar action is 
\eqn\accs{
S= \int d^5 x \sqrt { g_5} \bigg[
 { 1 \ov 2} \sum^3_{s=1} 
(\del_m \vp_s \del^m \vp_s + m^2_s \vp^2_s)
 + \l_{123} \vp_1 \vp_2 \vp_3 \bigg] \ , }
then  the corresponding 2- and 3-point correlators  of the boundary operators 
are  \freed\
\eqn\vxde{
 <O_{\D}(x_1)  O_{\D}(x_2)  >
 =    
 { a_0  \ov |x_1 -x_2|^{2\D} } \ ,\ \ \ \ \    \ \  a_0 =  
 { 2 (\D-2) \ov\pi^2  \D} {\G(\D + 1 ) \ov \G(\D-2)}\ ,   } 
\eqn\vde{
 <O_{\D_1}(x_1)  O_{\D_2}(x_2)  O_{\D_3}(x_3)>
 =  { \l_{123}\  a_1 \ov |x_1 -x_2|^{\D_1 + \D_2 - \D_3}
|x_1 -x_3|^{\D_1 + \D_3 - \D_2}
|x_3 -x_2|^{\D_3 + \D_2 - \D_1} } } 
 where 
$ \D_s = 2 + \sqrt{4 + m^2_s R^2}$  and\foot{Similar expression is found for 
any dimension $d$ of the boundary: the numbers $-2$ and $-4$ are replaced in general by
$-{d\ov 2}$ and $-d$.}  
\eqn\threepoint{
a_1 = -{\Ga [\ha (\D_1 + \D_2 - \D_3)] \Ga [\ha (\D_1 + \D_3 - \D_2)] 
\Ga [\ha (\D_2 + \D_3 - \D_1)] \Ga [\ha (\D_1 + \D_2 + \D_3 - 4)] \ov 2 \pi^4 
\Ga (\D_1 - { 2})\Ga (\D_2 - { 2})\Ga (\D_3 - {2})
}
}
It follows from \threepoint\ that whenever the dimensions of the operators $\Delta_s$ 
satisfy the relation (or any of its two permutations)
\eqn\diverg{
\D_3 = \D_1 + \D_2 + n\ , \,\,\,\,\,\,\,\,\,\,\,\,\,\, \ \ \ \ 
n=0,1,2, \dots \ ,
}
the coefficient $a_1$ develops a pole. This is precisely  what we encounter 
in  the case  of  the  $\p\p \nu$ interaction:
since $m^2_\p =0$ and $\ m^2_\n= {32\ov R^2}$ one has $\D_1=\D_2=4, \ \D_3 =8$, 
i.e. $\D_3 = \D_1 + \D_2$. 
The final result is thus {\it undefined}:
a product of a  zero and a divergence.\foot{If one regularizes
 the resulting IR  divergence 
(e.g., by shifting the position of the boundary)
the final result will  be zero because of 
the vanishing of the bulk interaction constant.} 

Our  proposal  is  that in such special cases one should use 
a special regularization prescription in 
computing  the supergravity  expressions. 
One should effectively  turn on ``a little bit'' of $S^5$ KK momentum
for the dilaton. Then 
(i) the dilaton  will get  an infinitesimal mass $\ep$, i.e. the 
dimension of $O_4$ will be  shifted from 4 to $4+\ep$, thus 
 regularizing \threepoint, and   
(ii) the 
 action will  get   a small coupling term  $\ep \p\p \nu$.
  As a result, the  dependence on $\ep$ will  cancel out 
in the final expression in \vde\   ($ \ep \times { 1 \ov \ep}$=1)   and 
we will  obtain a finite result  for $<O_4 O_4 O_8>$.
This procedure essentially amounts 
to an analytic continuation that treats the $S^5$-independent 
zero mode  $\p$  as an integral part 
of the whole  family of $S^5$ KK modes of the 10-d  dilaton field.

Equivalently, we  suggest that the AdS/CFT correspondence should 
be defined for the whole  KK families  of modes on the 
supergravity side, i.e.  one is to compute, e.g.,   the supergravity  $\p\p\n$ 
correlator  as a   limit of the correlator 
for   the two  massive  KK modes 
of the dilaton (with equal  masses $m^2_k= { k(k+4) 
\ov R^2}$, $k=0,1,2,...$)
and $\n$. A non-zero KK  quantum number $k$ 
implies  that now $\Delta_k = 4+k$  (the corresponding SYM operator is $O_{4+k} \sim
\Str (F^2 X^k  + ...))$  
 and thus the $\Gamma$-function factor in \vde\ 
$\Ga [\ha (\D_1 + \D_2 - \D_3)]$   is  simply 
$\Gamma(k)$. At the same time, the coupling constant $\l_{\phi\phi \nu}$
between $\n$ and the two KK dilaton  modes 
happens to be  non-vanishing for $k\not=0$ and  
is proportional to $k(k+4) \ov { 2^{k-1} (k+1) (k+2)} $. 
 As a result, the  product of $\l_{123}$ and $  a_1$ in \vde\
 is well-defined  and finite  ($ \sim k \Gamma(k) = \Gamma(1+k)$) 
    for {\it all $k$, 
  including the case of $k=0$}.

A similar prescription should be applied
 in all cases  where the anomalous dimensions
of the three operators in the correlation
function satisfy the relation \diverg. 
With a  hindsight, the suggested regularization
 procedure is actually implicit 
 in the examples discussed
in \sei, where  the 3-point functions
 $<\tr X^k \tr X^k \tr X^{2k}> \ (k=2,3,...)$
were   shown to  be  finite  and equal
 to their  large $N$ (free-field)  SYM 
counterparts. There
% Assuming analytic continuation in 
%$\Delta$'s (or masses), 
the {\it poles}  in  the supergravity amplitudes \vde,\threepoint\ 
were  similarly cancelled by {\it zeros}  in  the 
cubic Lagrangian coupling derived  from  the  
KK reduction  of  the 10-d supergravity action.\foot{The observation
about cancellation between poles and zeroes for  special 
combinations of  dimensions of the three operators in 
\sei\ was also  made  in   ref. \dhfr.}

Close parallel with the  case of the $\tr X^n$   3-point functions 
 and the supersymmetry relations  between the corresponding 
 operators on the SYM side \reew\
 leave little doubt that the  precise numerical AdS/CFT 
correspondence should then be found also in 
the present case of the normalized correlator 
$<O_4 O_4 O_8> \over {<O_4 O_4>^{1/2} <O_8 O_8>^{1/2} }$. 
To  check this  explicitly 
 one is to  compare  
the  finite numerical coefficient 
obtained using our prescription 
 on the supergravity side with the
 large $N$ SYM result. 
Since the comparison  is  done
  for the normalized 
correlator, one needs to know also the SYM expressions for $<O_4 O_4>$
and $<O_8 O_8>$. While   the SYM computation 
of $<O_4 O_4 O_8>$  (discussed below) 
 does not depend on the scalar 
and fermionic terms in the $\N=4$ 
supersymmetric completion of the $F^4$ operator
 \two,
 %\foot{The scalar and fermionic terms in the 
 %supersymmetric completion of $F^2$ operator  are proportional 
 %to equations of motion and thus may be ignored as such terms 
 %produce only contact terms in the correlators.  } 
  to find $<O_8 O_8>$ 
  one needs first  to determine  the full structure of $O_8$.
  This will be done  in \grey\ where we  will  
  present the full expression for the $\N=4$ super-invariant 
  $O_8$ in terms of $\N=1$ and $\N=2$ 
  superfields   and compute 
  $<O_8 O_8>$  using off-shell  superfield formalism.

%{\bf Expand the discussion here, need to explain why $F^4$ part 
%of $O_8$ is reliable and  outline the difficulty
%in getting a precise expression for scalar and fermion part.}
%\baselineskip 14pt

The plan of this  paper is   the  following.
In section 2 we shall present the SYM  calculation
for the correlator  $<O_4O_4 O_8>$ in the leading large $N$ (fixed $\l= \gym^2 N$) 
approximation.
In section 3
we shall determine  the supergravity expressions 
for the correlators $<O_4 O_4 O_8>$ 
using the  ``analytic continuation in KK momentum''
prescription outlined above. 
Section 4 will contain  some concluding remarks.
In Appendix A  we shall discuss an attempt  to determine
the normalization  of the $O_8$ operator  from its coupling to 
the fixed  scalar $\n$ in the Born-Infeld  D3-brane probe action.

%%%%%%%%%%%%%%%%%%%%%%%%%%%%%%%%%%%%%%%%%
\newsec{Gauge theory    calculation  of 
$<O_4  O_4 O_8 >$}
%%%%%%%%%%%%%%%%%%%%%%%%%%%%%%%

In this section we shall compute $<O_4  O_4 O_8 >$ correlator
in large $N$  SYM theory and demonstrate that it is non-vanishing.
As already mentioned above, 
%. Since we don't have a result for 
%$<O_8 O_8>$ yet, we are not able to produce a normalization 
%independent value of  $<O_4 (x) O_4(y) O_8(0) >$, which may be 
%compared to supergravity prediction.
%Thus for this paper, the result of this section  
%mainly serves as a demonstration that
%$<O_4 (x) O_4(y) O_8(0) >$ is non-zero.
one is able to find 
the precise expression for $<O_4  O_4 O_8 >$
(at separated points, i.e. modulo contact terms) 
without knowing the detailed  structure of
the supersymmetric completion of $F^4$ terms in \two.
It is 
% $O_8$ beyond those terms in \two\ is based on the observation that 
%the only the `irreducible' (terms which generate non-contact terms in
%correlation functions) terms in $O_4$ are those in \one.    
 easy  to understand 
   either  from the action of supercharges on 
the chiral primary  field \reew\ (see,  e.g.,  \ktv)
 or from the  ($\N \geq 2$)  superfield expressions \ferr\ 
that  the  scalar
and fermionic terms  in $O_4$ are of  the ``equations of motion"  
form\foot{Here we are ignoring
interaction (commutator $[X,X]$, $[\psi,X]$ dependent \ktv)  
terms in $O_4$
 since they produce terms of 
higher order in $\l$ in the correlators 
(which altogether should cancel out in the cases where there is a non-renormalization theorem). Since the dilaton should be related
to  gauge coupling constant, one expects that the dilaton operator should 
be proportional to the SYM Lagrangian,  up to a total derivative term.
The $\N=2$ supersymmetry implies that the quadratic scalar  term
should have the form $\tr ( X D^2 X)$, which is different from the 
standard kinetic term by a total derivative. }
\eqn\trw{O_4 =  \tr (  F^2 +  X  \del^2 X + \bar \psi \gamma \cdot \del \psi+ ... ) \ . }
Contracted with other fields in the correlators 
these terms  give rise only to contact contributions 
 proportional to delta functions 
 (which are field-redefinition dependent and  are 
 omitted in CFT correlation functions).\foot{This explains, in particular, 
the  agreement between the result obtained 
using the dilatonic $F^2$  operator that
 follows from the BI action 
\refs{\kleb,\klebb} (in the Einstein frame
the BI action  gives $e^{-\p}  F^2 +  (\del X)^2 + ...$,
i.e. scalars and fermions do not couple to $\p$)
and the form of the 
 corresponding  $\N=2$ 
superinvariant   that has the structure \trw.    
The scalar and fermionic terms are important, however, 
in  most of other relevant  cases: for example, 
the stress tensor operator that couples to the  4-d graviton  contains
non-trivial scalar and fermionic terms \klebb\ (see also \ktv).}

As a result,  the   scalar  and fermionic terms in  $O_4$ and 
$O_8$ are irrelevant for the computation of $<O_4(x) O_4(0)>$ and 
$<O_4 (x) O_4(y) O_8(0) > $ which 
 can therefore be done  simply in the pure YM theory.
The  correlator  $<O_8(x) O_8(0)>$  does, however, receive non-trivial 
scalar and spinor contributions, i.e.  is  
different from the pure YM result \grey.

We shall  take the $SU(N)$ Yang-Mills Lagrangian in the form 
\eqn\laa{
{\cal L} = {N \ov 4\l } \tr (F_{\m \n} F_{\m \n}) \ ,
}
where $\l = \gym^2 N$.  We are assuming 
 that the generators
are normalized so that $\tr (T_i T_j) = \d_{ij}$.\foot{A more
 conventional  choice of normalization
of the generators in the fundamental representation is 
 $\tr (T_i T_j) = \ha \d_{ij}$, 
% That effectively redefines the gauge coupling. In this case  $\l = 4 \pi %g_s N $
\  ${\cal L} = {N \ov 2\l } \tr (F_{\m \n} F_{\m \n})$.}
In the double-line notation, $A^a_{\m b} = A_\m^i (T_i)^a_b$, the 
gauge vector  propagator is 
\eqn\gluon{
<A^a_{\m b}(x) A^c_{\n d}(0)> =  {\l \ov N} (\d^a_d \d^c_b - {1 \ov N}
\d_b^a \d^c_d){  \d_{\m \n}\ov 4 \pi^2 x^2} \ . 
}
In the free-theory limit the field strength propagator is then
\eqn\ff{
<F^a_{\m \n b} (x) F^c_{\l \r d} (0)> =  
{\l \ov N} (\d^a_d \d^c_b - {1 \ov N} \d_b^a \d^c_d)
{2 \ov \pi^2 x^4} D_{\m \n , \l \r}(x) \ , 
}
where 
\eqn\tend{
D_{\m \n , \l \r}(x) =  \ep_{\m \n , \l \r} - {4  \ov x^2}
X_{\m \n , \l \r}\ ,
}
$$
\ep_{\m \n , \l \r}  \equiv \ha (\d_{\m \l} \d_{\n \r} - \d_{\m \r} \d_{\n \l}) \ , \ \ \ \ \ \ \  
X_{\m \n , \l \r}  \equiv {1 \ov 4} (x_\m x_\l \d_{\n \r} - x_\m x_\r \d_{\n \l} 
+ x_\n x_\r \d_{\m \l} - x_\n x_\l \d_{\m \r}) \ . 
$$
All   correlators  will be  found  in  the free theory
limit, i.e.  ignoring  all contributions
which are sub-leading at large $N$  and small  $\l$. Hence 
 the field strength matrices in \one\ and \two\ 
do not contain the interaction term, i.e. are  simply
 $F_{\m\n}= \del_\m A_\n - \del_\n A_\m$,
and the computation of the correlators 
reduced to a  straightforward application of \ff.
After   lengthy
calculations  
one finds\foot{In the case of 
{ abelian}   gauge theory these correlators were computed 
%(also ignoring  fermionic and scalar terms in $O_8$) 
in 
\kle. The numerical coefficients in the correlators found
in the abelian theory 
and the large $N$ non-abelian theory are  {\it different}. }
\eqn\ffs{
<O_4(x) O_4(0) > = {48 \l^2\ov \pi^4} {1 \ov x^8} (1 - {1 \ov N^2})\ ,
}
%\eqn\fffs{
%<O_8(x) O_8(0) >_{\rm g.f.} 
%=  {8 \times 27 \l^4 \ov \pi^8} {1\ov x^{16}} + O({1\ov N^2})\ , 
%}
\eqn\ooof{
<O_4 (x) O_4(y) O_8(0) > 
= {1 \ov N} {128 \times 9 \l^4\ov \pi^8} 
{ 1\ov x^8 y^8} + O({1\ov N^3})\ . 
}
Similarly, one finds that the gauge-field part   
${ 3 \ov 2} \Str[ F^4-\four (F^2)^2] $
 of the operator $O_8$  contribution to the 
correlator $<O_8(x) O_8(0) >$ is 
\eqn\fffs{
<O_8(x) O_8(0) >_{\rm g.f.} 
=  {8 \times 27 \l^4 \ov \pi^8} {1\ov x^{16}} + O({1\ov N^2})\ . 
}
The contributions of the 
fermionic and scalar terms in $O_8$  increase  the numerical
coefficient in \fffs\ \grey.

%%%%%%%%%%%%%%%%%%%%%%%%%%%%%%%%%%%%%%%%%
\newsec{$D=5$ supergravity  results for the correlators}
%%%%%%%%%%%%%%%%%%%%%%%%%%%%%%%

The type IIB supergravity action 
for  the (Einstein-frame) metric,
 dilaton,  RR scalar  and 4-form  is ($ 2 \k_{10}^2  = { (2 \pi)^7 g_s^2 \a'^4}$)
\eqn\tena{
I_{10} = -{1 \ov 2 \k_{10}^2 } \int \! d^{10} x \sqrt{g} \, 
\bigg[ {\cal R} - \ha (\del \p)^2  - \ha e^{ 2 \p} (\del C)^2 - { 1 \ov 4\cdot  5!} (F_5)^2  \bigg] \ . 
}
In what follows we shall concentrate on the dilaton -- fixed scalar
correlators, but the case where the  dilaton is replaced by the RR scalar $C$
(and  the SYM operator $\tr F^2$ is replaced by $\tr FF^*$) 
is  very similar.
The  background $AdS_5 \times S^5$  metric  (which is an extremum  
of \tena\ \schwa) is 
\eqn\bac{
ds_{10}^2 = R^2 [{1 \ov z^2} (dz^2 + dx_\m  dx_\m ) + d \Omega^2_5] 
\ , \ \ \ \ \ \ \ \ \     R^4= 4 \pi g_s N \a'^2 \ . } 
In what follows we shall  set the radius $R$ to be 1. 
To dimensionally reduce to $D=5$, we use  the ansatz \radi\ for the 
 10-d Einstein frame
metric.  
%\eqn\meet{
%ds^2_{10} = e^{-{10 \ov 3} \n(x)} g_{5 mn} (x) dx^m dx^n + 
%e^{2 \n(x)} d \Omega_5^2 \ . }
The  relevant part of the type IIB
supergravity action reduced to 5 dimensions 
for the zero modes of the dilaton   and  the metric is 
 \eqn\action{
I_5 = - {1 \ov 2 \k_5^2} \int \! d^5 x \sqrt{g_5} \, 
\bigg[ {\cal R}_5 - \ha (\del \phi)^2  - {40 \ov 3} (\del \n)^2 -  V(\n) 
  + ... \bigg]  \ , 
}
where\foot{The two terms in the potential originate from
 the curvature of $S^5$ 
and the $(F_5)^2$  term.}
\eqn\poot{
V(\n) = 8 e^{-{40 \ov 3} \n} - 20 e^{-{16 \ov 3} \n} =
- 12 + { {40 } \ov 3}\times 32 \n^2 +  O(\n^3) \ . }
 This action does {\it not} contain $\p$--$\n$ coupling 
(in particular, $\phi\phi\nu$ term) -- this is a consequence of the fact that 
   the dilaton does not couple to both $\cal R$ and $(F_5)^2$ in \tena.

As was already explained in the Introduction, 
our  main  suggestion is that  to  establish a  correspondence 
between the SYM and  the supergravity  expressions for the 
 $<O_4O_4O_8>$ correlator 
one should ``regularize'' the dimensional reduction procedure. 
One should first assume that $\p$ has some  ``small'' dependence 
on $S^5$ coordinates, leading to  an  infinitesimal 5-d dilaton
  mass term and an  infinitesimal $\p\p\n$ coupling, 
and take   the corresponding small parameter 
to zero only after the computation of the 
$\p\p\n$ correlator.
% (which will turn out to be finite in this limit).
%This IR regularization  procedure is somewhat analogous 
%to analytic continuation of some  flat space  string amplitudes 
%in external momenta.
The meaning  
of this prescription  is that  treating   
 the zero mode of the  dilaton on the same footing as higher  KK  modes 
  resolves  the ambiguity in the definition  of the 
$\p\p\n$ correlator and leads to a finite value for its 
overall numerical coefficient.

Let us start 
with  decomposing  $\p$ in terms of spherical harmonics on $S^5$  
\eqn\deco{
\p(y,x)  = \sum_I  Y^I (y)  \p^I(x) \ . }
%where $\g_s$ are  the coordinates of $S^5$. 
We shall  follow the notation of \refs{\sei}  and  consider $S^5$
as embedded in  a Euclidean space $R^6$. The spherical harmonics  $Y^I$ are defined
by  $Y^I = C^I_{i_1 ... i_k} y^{i_1} \cdots y^{i_k}$, where $y^i$ ($i=1,...,6$) 
are the 
Cartesian coordinates of the embedding  $R^6$ space ($y^iy^i=1$)    and $C^I_{i_1
... i_k}$ are  totally symmetric traceless tensors of rank $k$.   

The  $I=0$ term  in \deco\  $\p^0(x)\equiv \p(x) $ 
is the 5-dimensional dilaton while $\p^I(x)$ are higher KK modes 
of the $D=10$  dilaton.
The dimensionally reduced form of the 
 dilaton  part of the action \tena\ is  found to be
\eqn\dimre{
{1 \ov 2 \k_{10}^2 } \int \! d^{10} x \  \ha \sqrt{g}\ 
g^{MN} \del_{M} \p 
\del_{N} \p  = {1 \ov 2 \k_{5}^2 }
\int \! d^{5} x \sqrt{g_{5}} (L_1 + L_2) \ , 
}
where ($\om_5=\pi^3$)
\eqn\kka{
{1 \ov 2 \k_5^2} = {R^8 \om_5  \ov 2 \k_{10}^2 } 
= {N^2 \ov 8 \pi^2}
\ .  }
%and   is the volume of a unit 5-sphere.
The two parts of the dilaton Lagrangian are 
\eqn\tyr{
L_1 = \ha g_{5}^{m n} \sum_{I} A(k)  \del_{m} \p^I 
\del_{n} \p^I
\ , } 
and 
\eqn\tiuy{
L_2 = \ha e^{-{16 \ov 3} \n} \sum_{I} B(k) \p^I  \p^I
\ , } 
with  the normalization constants $A(k)$ and $B(k)$ given by
($k$ is the rank of $C^I_{i_1 ... i_k}$)
\eqn\nji{
A(k) = {1 \ov 2^{k-1} (k+1) (k+2)}\ , \ \ \ \ \ \ \ 
B(k) = k(k+4) A(k) \ .
}
The  factors of  $e^\n$ in \radi\ where chosen 
so that the Einstein ${\cal R}$-term and  hence  $L_1$ have no $\n$-dependence.
The form of the  $\n$-dependent factor in $L_2$ \tiuy\
follows  from the fact that  
$L_2$ =  $e^{-{10 \ov 3} \n(x)} \int_{S^5} g^{ss'} \del_s \p \del_{s'} \p,$
where $g_{ss'}\sim e^{2\n}$ is the $S^5$ part of the metric  in 
\radi. 

Since for $I=0$  one has $k=0$  and thus vanishing $B(k)$, 
we conclude once again 
that there is no coupling between $\n$ and the zero mode 
of the  dilaton. 
Instead of taking  $k=0$ directly, 
one may formally 
set   $k= \ep$   and take the limit
 $\ep\to 0$  only at the end of the calculation 
of correlation functions.
As follows from \tiuy,
 this  prescription 
%of turning  on `a little bit of   KK momentum'
 generates { both} a small mass  for the 5-d dilaton 
{ and}  a coupling between  the dilaton  and
$\n$,
\eqn\yoo{
m^2 = 4 \ep\ , \ \ \ \ \ \ \ \ 
\ \ \ 
\l_{\p \p \n} =  -{32 \ov 3} \ep= -{8 \ov 3} m^2  
\ . } 
%Let us note that similar conclusion  
%is found by using a different (more artificial) prescription
%of starting with a small dilaton mass term  $m^2 \p^2$ 
%added directly to  the $D=10$
%action \tena. Then the $D=5$ action will contain
%$ {1 \ov 2 \k_5^2} \int \! d^5 x \sqrt{g_5} \, 
%\big[  \ha (\del \phi)^2 +
%\ha e^{-{10 \ov 3} \n } m^2 \p^2\big] $.
%  Thus the cubic  $\p\p\n$ coupling  will be again   proportional to $m^2$, 
%though the numerical coefficient is obviously different, 
%$\l_{\p \p \n} = - {5 \ov 3} m^2 $,
%reflecting the fact that here the mass term 
%has artificial 10-d and not 5-d KK origin. 
Introducing  the couplings between the $AdS_5$
boundary values of the 5-d   fields $\p$ and $ \n$ and
the corresponding  $\N=4$  SYM 
$F^2$ and $F^4$  operators \one,\two\  as in \refs{\gkp,\witt} we get 
\eqn\inter{
\int \! d^4 x \, \big[ \ a_4\  O_4\  \p(x)  + \ a_8\  O_8\  \n(x)\big]   \ ,
}
where $a_4$ and $a_8$ are some  normalization constants.  

We learn from 
  \action\ and \inter\  that the  operators 
 $O_4$ and $O_8$  have the following dimensions (${m^2 \to 0}$)
\eqn\diim{
\D_4 = 2 + \sqrt{4 + m^2} \ = \    4 + {m^2 \ov 4} + O(m^4)\ , 
\ \ \ \ \ \ \ \ 
\D_8 = 2  + \sqrt{4 + 32}\  =\  8 \ . }
%Thus the role of our regularization is to shift the 
%dimension of $O_4$ away from the special point 4. 
Following \refs{\gkp,\witt} and  using the expressions 
for the 2- and 3-point functions in \freed\ 
we find from \action,\inter\
the following supergravity predictions
for the correlators involving $O_4$ and $O_8$ (see eqs. \vxde--\threepoint)
\eqn\fff{
<O_4 (x) O_4(0) > = {1 \ov 2 \k_5^2 }   {24 \ov \pi^2  a_4^2 } { 1 \ov x^8}
\ , }
\eqn\ffff{
<O_8(x) O_8(0)> =  {1 \ov 2 \k_5^2}  
{42 \times 24 \times 40 \ov 3 \pi^2 a_8^2} {1 \ov x^{16}}
\ , } 
\eqn\ooos{
<O_4(x) O_4(y) O_8(0)> =  {1 \ov 2 \k_5^2} 
{8 \times 36  m^2  \ov 3 \pi^4 {a_4^2 a_8}} 
  \Ga ({m^2 \ov 4}) { 1 \ov x^8 y^8}\ 
=    {1 \ov 2 \k_5^2} 
{32\times 36  \ov 3 \pi^4 a_4^2 a_8} { 1\ov x^8 y^8} + O(m^2) \ . 
}
There is  an extra factor  of 2  in \ooos\ coming from the fact   that there are 
two dilaton field legs  in  the $\p\p\n$ coupling 
\yoo. In \ooos\ we  have used that  for 
 $m^2 \to 0$, \  $\Ga({m^2\ov 4}) \to  {4 \ov m^2}$.
The resulting expression for 
the $\p\p\n$  correlation function 
 is thus {\it finite}:
 %  even though the $\p\p\n$ coupling  in the action vanishes for $m^2=0$:
the  zero in the $\p\p\n$ bulk interaction vertex cancels against
the pole in the $\Gamma$-function
coming from the integral over $AdS_5$.

% which was effectively what 
%happened in \sei. 

Let us note in passing  that 
there is  also no  $\p \n\n$ coupling in \action, i.e.
\eqn\poos{
<O_4(x) O_8(y) O_8(0)> = 0 \ . }
Since the dimensions of the three  operators here do not satisfy \diverg,
 in contrast to  the case of $<O_4(x) O_4(y) O_8(0)>$
the expression \threepoint\ for  this 
correlator  is  well-defined and there is no ambiguity.
The vanishing of 
$<O_4(x) O_8(y) O_8(0)>$  %(but not  $<O_8(x) O_8(y) O_8(0)>$)
in \poos\ can be deduced from the argument based on 
$U(1)_Y$ charge  conservation, see \refs{\intri,\eden}. From the 
supergravity point of view, this $U(1)$ is part of $SL(2,R)$
which is a classical symmetry of type IIB supergravity.
The dilaton transforms non-trivially under this group, 
while $\n$ is inert being part of the Einstein-frame metric.
Thus the coupling like $\p \n\n$ is forbidden in the classical 
supergravity Lagrangian.

%Correspondingly, in the SYM theory, 
%one can assign the following values of the $U(1)$
%charge to the relevant  operators:
%$$O_4^{(\pm)}:  \ q=\pm 4\ ,\  \  \ \ \  \ \ \ O_8:\  q=0\ ,  $$
%where $O_4^{(\pm)} \sim (F^{(\pm)})^2 + ..., \ 
%F^{(\pm)} = F \pm F^*$. 
%As a result, one can argue 
% using the $\N=4$ superspace 
% formalism  \refs{\howes,\intri,\eden}
%that while  $<O_4^{(+)} O_4^{(-)} O_8>$  and $<O_8 O_8 O_8>$
%may be non-vanishing, 
%$<O_4^{(\pm)} O_8 O_8>$ must vanish since the total $U(1)$ 
%charge here  is non-zero 
%while  it must be conserved at the level of 3-point functions. 
%This is in obvious correspondence with what 
%one finds from the supergravity action.

%    (finite)
%when   computed  by first introducing 
% some $\p \n\n$ vertex into the action.

%To test the AdS/CFT correspondence  we are now to 
%compute the correlators in \fff, \ffff, \ooos, \poos\
%directly in  the  large $N$   SYM theory at weak coupling
%We shall use the free-theory  approximation
%since it is natural to expect that the 2- and 3-point
%correlators of chiral primary operators 
%are not renormalized. 
%Then  the check 
%of the AdS/CFT correspondence should be the agreement
%between the  supergravity (strong-coupling  SYM)
%and weak-coupling SYM expressions.

Since the supergravity expressions \fff--\ooos\ depend on the 
normalization constants appearing in \inter,  the  direct 
comparison with SYM result \ooof\ is not possible. The weighted 
three-point function does not depend on the absolute normalization,
\eqn\wthre{
<\OO_4(x) \OO_4(y) \OO_8(0)> = {1 \ov N} \sqrt{{16 \ov 105}} {1 \ov x^8 y^8}
\ , 
}
where $\OO_4$ and $\OO_8$ are rescaled $O_4$ and $O_8$  satisfying  
\eqn\normalize{
<\OO_4(x) \OO_4(0)> = {1 \ov x^8}\ , \ \ \ \ \ \ \  
<\OO_8(x) \OO_8(0)> = {1 \ov x^{16}} \  .
}
To compare \wthre\ with  the SYM result we need  
the  expression for  $<O_8 O_8>$ in SYM  theory  which 
requires a detailed  separate computation 
and will   be presented   in \grey.
Here we note only   that the value of  $<O_8 O_8>$ in SYM theory 
which would lead to the  agreement with \wthre\ is
\eqn\ffe{
<O_8(x) O_8(0) > =  {105 \times 36  \ov \pi^8} {\l^4 \ov x^{16}} \ . }
Not surprisingly,  this  expression is different 
 from the pure gauge theory 
part \fffs\ of the full SYM correlator.

One heuristic  way   to fix the  values of the normalization constants  
 $a_4$ and $a_8$ in \inter\ is to derive \inter\
from the BI action  for a D3-brane  propagating in a curved background.
As shown in Appendix  this leads to 
\eqn\coeff{
a_4 = {N \ov 4\l}\ , \ \ \ \ \ \  \ 
a_8 =  {10 N \pi^2    \ov 9\l^2} 
\ , }
where $\l= \gym^2 N = 4 \pi g_s N$. 
The leading large $N$ term in \ffs\ is in perfect agreement 
with the supergravity expression \fff\  with $a_4$ given by
the BI normalization in \coeff\ (this agreement
is essentially  the one  originally found 
in \refs{\kleb,\klebb,\gkp}). 
If we use $a_8$ from \coeff\ in \ffff\ we 
find that the supergravity expression  for $<O_8(x) O_8(0) >$ is 
\eqn\suuu{
<O_8(x) O_8(0) >_{\rm BI}
 =  {84 \times 81 \l^4\ov 5\pi^8} {1 \ov x^{16}} \  , }
and from \ooos\ we get 
\eqn\suuv{
<O_4 (x) O_4(y) O_8(0) >_{\rm BI} 
= {1 \ov N} {96 \times 36 \l^4\ov  5 \pi^8} { 1\ov x^8 y^8}
\   .  }
The numerical  coefficient  in \suuv\ is different from
the one in  \ooof\  by factor of $3 \ov 5$. 
This disagreement may  be related to  ambiguities 
%seems to suggests  that one cannot 
in using the BI action for fixing the normalization
of  operators beyond the leading second-derivative 
terms (see Appendix).

%The reason for the  disagreement 
%may be  the  use of the heuristic procedure based 
%on Born-Infeld  action to fix
%the  normalization  constant $a_8$ in \coeff (see discussion in 
%appendix).

%%%%%%%%%%%%%%%%%%%%%%%%%%%%%%%%
\newsec{Concluding remarks}
%%%%%%%%%%%%%%%%%%%%%%%%%%%%%

What follows from our  proposal is 
that though some of the couplings  in the  $S^5$  compactified  5-d 
(supergravity + KK multiplets)
 action  may  appear  as  missing they are actually `hidden' 
and must be taken into account  in establishing the 
 AdS/CFT
correspondence.
This  may  
  carry an important message  about how 
the AdS/CFT correspondence is actually working.
The  10-d picture with non-zero KK
momentum  seems to  be  more fundamental than the 5-d one.
This is after all  what the original precise conjecture \mal\
about the  duality of {\it ten-dimensional} 
  string on $AdS_5 \times S^5$  and  $\N=4$ SYM 
theory is implying.
% (as compared to 
%broader  notions of  holography and   AdS/CFT correspondence \witt). 

On the supergravity/string side the 5-d fields 
appear as KK modes of $D=10$ fields upon compactification on $S^5$.
The families of 5-d fields  originating 
from a single type IIB supergravity   field should definitely `remember'
 their common 10-dimensional  origin. At the same time, 
this  10-dimensional  origin of the corresponding composite operators 
on the SYM side is quite obscure. The $AdS_5\times S^5$ string -- SYM 
correspondence suggests that there  may exist 
a `10-dimensional' reformulation of $\N=4$ SYM  theory 
where these  operators are  somehow 
unified in `KK families'. 
%It might be that this 
%`10-dimensional' reformulation may also make the
%maximal   supersymmetry of the theory more explicit.
%This may also reveal some  hidden symmetries of the SYM theory 
%whose existence is  related to the $D=10$ origin of the
%dual $AdS_5 \times S^5$ string theory.

One of our motivations for  trying to get a 
detailed understanding of how the  AdS/CFT correspondence 
works at the level of 3-point functions  is 
to shed some light on  the factorization properties 
of  much more complicated 4-point correlators.
%Our discussion should be also relevant
%for the computation of four-point functions
%of $\N=4$ SYM theory.
 In the previous discussions of the supergravity
expression for the 4-point function
of the `dilatonic'  $O_4$   operator 
 \refs{\LT, \fref, \chal,\fres}, the 
contribution of the scalar 
 $\nu$  exchange to the  four-dilaton 
 amplitude  was  not included.
This  may  be   justified  by noting that 
 although the 3-point amplitudes \vde\  are divergent  
for the  operators satisfying \diverg, the  exchange 
amplitude  contributing to the 4-point function 
 $<O_{\D_1} O_{ \D_1} O_{\D_2}  O_{ \D_2} >$
 with  the exchanged state corresponding to 
the operator $O_{\D_3}$ 
%as an intermediate state
 is actually finite \refs{\liu,\dhfr}. 
Naively, one   may   conclude that
 the $\n$-exchange contribution  to  the  four-dilaton 
 amplitude is  indeed zero. 
%Note, however, that 
This  seems to  provide an example 
when the existence of a  non-vanishing  three-point
function ($<O_4O_4O_8>$)
 does not imply the presence of the corresponding factorization
contribution to  the 4-point ($<O_4O_4O_4O_4>$)
 correlator.
% in conflict with  naive expectations based on
% OPE for a closed set of operators.\foot{There may exist
%examples  of CFT's   (with continuos  spectrum, etc.) 
%in which  such naive expectations may not actually apply.}
%one of the operators in the
%OPE of the other two. 
%It is likely, however, 

The  story  may  be more intricate.
In line with our  proposal  for the   3-point function $<O_4O_4O_8>$
we suggest that the  supergravity expression 
for the 4-dilaton amplitude  should be defined as a $k\to 0$   limit 
of the amplitude $<O_{4+k}O_{4+k}O_{4+k}O_{4+k}>$ 
for four  massive ($m^2_k = {k(k+4)\ov R^2})$ 
 dilaton modes. 
 This  amplitude receives contributions 
from  both  graviton and  fixed scalar exchanges.
 It may happen that  when the  correlator $<O_4O_4O_4O_4>$  is 
defined in this way, its  factorization 
will become more transparent (cf. \refs{\fref,\liu,\dhfr,\fres}).

\newsec{Acknowledgements}
%%%%%%%%%%%%%%%%%%%%%%%%%%%%%
We are   grateful to 
%I.  Buchbinder,  B. Eden,
 F. Gonzalez-Rey, D. Gross, A. Hashimoto,  G. Horowitz, 
%P.  Howe,  
I. Klebanov,   J. Polchinski
% M. Ro\v cek   
and E. D'Hoker for useful discussions
and comments.  
%at different   stages of this work.
%for  useful  discussions 
This work was supported  by PPARC,  the European
Commission TMR grant  ERBFMRX-CT96-0045,  
INTAS grant No.96-538,  
 Nato 
 %Collaborative Linkage 
 Grant PST.CLG 974965
   and NSF grant PHY94-07194.
%%%%%%%%%%%%%%%%%%%%%%%%%%%%%%%%%%%%%%%%%%%%%%%%%%%%%%%%%%%%%%%%%%%%%

%%%%%%%%%%%%%%%%%%%%%%%%%%%%%%%%%%%%%%%%%%%%%%%%%%%%%%%%%
\appendix{A}{Normalizations  of $O_4$ and $O_8$
operators  from Born-Infeld  action}
%%%%%%%%%%%%%%%%%%%%%%%%%%%%%%%%%%%%%%%%%%%%%%%%%%%%%%%%%

Here we describe  an attempt  to fix normalizations 
in \inter\ by starting with a Born-Infeld action.\foot{Similar 
approach of fixing normalizations from BI action was used in 
\refs{\klebb,\dastr}.}
The Born-Infeld Lagrangian for (a collection of)  D3-branes   in
curved   background has the form:
\eqn\bia{
{\cal L} = T_3  \ 
{\rm Str} \sqrt{\det ({\hat g_{\m\n}} + e^{-{\p \ov 2} } \cF_{\m\n})}
+ \  {\rm WZ-terms} \ , 
}
where 
\eqn\defs{
\cF_{\m\n}  = 2 \pi \a'  F_{\m\n} \ , \ \ \ \  \ \  \ \ \ \
T_3 = {1 \ov (2 \pi)^3 g_s \a'^2 }\ . } 
We take the background metric Einstein-frame 
metric  as a combination of \bac\ and \radi:
\eqn\bgme{
ds^2 = R^2 [e^{-{10 \ov 3} \n} {dz^2 + dx_{\m}  dx_\m \ov z^2} 
+ e^{2 \n} d \Omega_5^2 ]\ ,  
}
\eqn\yyt{
R^4 =  \l \a'^2 \ , \ \ \ \ \   \l= 4 \pi g_s N = \gym^2 N \ . }
Then the relevant part of the Lagrangian 
in \bia\ becomes
\eqn\biex{\eqalign{
{\cal L}  & = T_3 ({R^2 \ov z^2} e^{-{10 \ov 3} \n})^2  
{\rm Str} \bigg(\bigg[\det (\d_{\m \n} 
+ e^{-{\p \ov 2} } {z^2 \ov R^2}
 e^{{10 \ov 3} \n} \cF_{\m\n})\bigg]^{\ha} 
- 1 \bigg) \cr
& =  T_3 ({R^2 \ov z^2} e^{-{10 \ov 3} \n})^2 
{\rm Str} \bigg[\   {1 \ov 4} e^{-\p} ({z^2 \ov R^2} e^{{10 \ov 3} \n})^2 \cF^2 
- {1 \ov 8} e^{- 2 \p} ({z^2 \ov R^2} e^{{10 \ov 3} \n})^4 
\bigg(\cF^4 - {1 \ov 4} (\cF^2)^2 \bigg) + ...\bigg] \cr
& =   
 {1 \ov 4} T_3 e^{-\p} (2 \pi \a')^2 O_4 
- {1 \ov 8} T_3 e^{- 2 \p} ({z^2 \ov R^2} e^{{10 \ov 3} \n})^2
(2 \pi \a')^4 {2 \ov 3} O_8  + ...\cr
& =   ...
 - \bigg( {N \ov 4\l} \p O_4 
+  {10 N \pi^2 \ov 9\l^2} \n z^4 O_8\bigg) 
    +  {N \pi^2 \ov 3\l^2} \p z^4 O_8
 + ... \ . 
}}
Here we used the definitions of $O_4$ and $O_8$ in \one,\two.
Comparison with \inter\ then implies\foot{In taking the boundary limit
the   factor $z^4$  produces 
 a cutoff factor which should multiply 
$F^4$ on dimensional grounds.}
\eqn\coom{
a_4 ={N \ov 4\l}\ , \ \ \ \ \ \ \ \ \ \  
a_8 =  {10 N \pi^2 \ov 9\l^2}
\ .  }
The normalization of $O_8$ derived in this way is  
only heuristic and should be treated with caution. 

Indeed, this argument may look suspicious for several reasons. 
First,  $\n$ already couples to scalars $X$ at the  second derivative
level but   $(\del X)^2$ cannot  be  part of the $O_8$ 
operator. Also,   $\n$ interacts {\it differently}  with 
$F^4$ and $(\del X)^4$  terms,   in contradiction with the fact that 
they must be parts of the same super-invariant.
The scalar partners of  the gauge field couple to $\n$ at quartic order
as $ e^{4\n} (\del X)^4$, which leads to the  normalization coefficient   
different from the one in  \coom\ -- since  the coefficient of $\nu$ 
 $10 \ov 3$  is  replaced by $ 2$,  we get  
\eqn\comm{
a'_8 =  {2N \pi^2 \ov 3\l^2}
\ .  }
Remarkably, this is {\it precisely}  the coefficient one needs for  the 
correspondence between
the supergravity and SYM expressions for the 
3-point functions (see eqs.  \ooos\ and \ooof). 
However, due to  the ambiguities mentioned above it seems  hard 
at the moment to make 
this argument convincing.

In general, the 
BI action
 \refs{\frad,\napp,\leigh}
summarizes the  low-energy limit of  scattering amplitudes
of  massless open string   modes
in flat space.  Its direct   curved-space 
generalization correctly accounts for  certain zero-momentum
 limits of 
the  disc amplitudes  with    few  closed string 
vertex operator insertions (see, e.g., \disc). 
One should keep in mind, however, 
that the standard BI action  does  not include terms with 
derivatives  of  the vector field strength,  higher than 
the first derivative
of the scalar (coordinate) fields
and  terms with derivatives  of the closed string fields.
As a result, there may be a clash with manifest
(linearly realized) $\N\leq 4$ supersymmetry:
the   manifestly supersymmetric 
extensions of the    structures  appearing 
in the expansion of the  bosonic BI action  may involve 
higher derivatives; elimination of such higher-derivative
 terms via integration by parts 
may  produce terms  with derivatives of the 
background  supergravity fields, i.e. terms  which are 
not included  
in the standard BI  action.
In particular, one should not be surprised if one 
does not find  the correct bosonic (scalar) parts of the 
corresponding superinvariants   from the expansion of the BI action.

%We may look at how the scalar partners of  the gauge field couple to $\n$.
%These are, in particular,  fluctuations 
%in $S^5$ directions (we assume that brane probe is put at a fixed 
%radius $z$). They   will enter \biex\ 
%as $ e^{4\n} (\del X)^4$, and that will indeed lead 
%to the  normalization coefficient   different from the one in  \coom: 
%with $10 \ov 3$ replaced by $ 2$, i.e. 

%May be we need 4+6 split;
%This BI action is SYM eff action  with $z$ as an IR cutoff;
%how it is to be coupled to $\n$??
%In \kle\ they were thinking of 4d metric $h_{\m\m}$ as a fixed scalar.

%So far truth is that $\n$ interacts $differently$ with 
%$F^4$ and $(\del X)^4$   terms ...
%in contradiction with  the fact that they must be parts 
%of the same super-invariant.

%%%%%%%%%%%%%%%%%%%%%%%%%%%%%%%%%%%%%%%%%%%%%%%%%%

\listrefs
\bye